\documentclass[12pt,a4paper]{article}
\usepackage{graphicx}
\usepackage{dcolumn}
\usepackage{bm}
\usepackage{amsmath,amsfonts,amssymb,mathrsfs}
\usepackage[dvipsnames]{xcolor}
\usepackage{bbm}
\usepackage{slashed, MnSymbol}
\usepackage{braket,xcolor}
\usepackage{verbatim}
\usepackage{subcaption}
\usepackage[normalem]{ulem}
\usepackage{multirow}
\usepackage{amsfonts,mathtools,resizegather}
\usepackage[normalem]{ulem}
\usepackage[utf8]{inputenc}
\usepackage{scalerel}
\usepackage{makecell}
\usepackage{dcolumn}
\usepackage{amsfonts, mathtools,resizegather}
\usepackage{multicol}
\usepackage{subcaption}
\usepackage[dvipsnames]{xcolor}
\usepackage[bb=boondox]{mathalfa}
\usepackage[export]{adjustbox}
\usepackage[normalem]{ulem}
\usepackage{caption,jheppub}

\usepackage{xcolor}
\usepackage{bm}
\usepackage{graphicx, physics}
\usepackage{amsmath}
\usepackage{amssymb}
\usepackage{amsbsy}
\usepackage{amsthm}
\usepackage{amsfonts}
\usepackage{upgreek}
\usepackage{bbm}
\usepackage{subcaption}
\usepackage{amsfonts}
\usepackage{slashed}
\usepackage{jheppub}
\usepackage{algorithm}
\usepackage{algpseudocode}

\title{\Large Tridiagonal Hamiltonians modeling the density of states of the Double-Scaled SYK model}

\author[a,b]{Pratik Nandy\,\href{https://orcid.org/0000-0001-5383-2458}
{\includegraphics[scale=0.05]{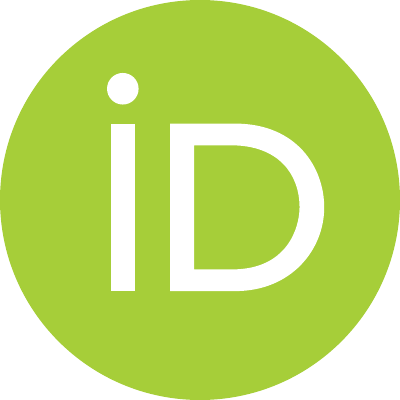}}\,}

\emailAdd{pratik@yukawa.kyoto-u.ac.jp}

\affiliation[a]{Center for Gravitational Physics and Quantum Information,\\
Yukawa Institute for Theoretical Physics, Kyoto University,\\
Kitashirakawa Oiwakecho, Sakyo-ku, Kyoto 606-8502, Japan}
\affiliation[b]{RIKEN Interdisciplinary Theoretical and Mathematical Sciences Program (iTHEMS),\\
Wako, Saitama 351-0198, Japan}

\abstract{By analyzing the global density of states (DOS) in the Double-Scaled Sachdev-Ye-Kitaev (DSSYK) model, we construct a finite-dimensional Hamiltonian that replicates this DOS. We then tridiagonalize the Hamiltonian to determine the mean Lanczos coefficients within the parameter range. The bulk Lanczos coefficients, especially the \emph{Lanczos descent} can be analytically expressed as a particular $q$-deformation of the logarithm. Our numerical results are further corroborated by semi-analytical findings, a random matrix potential construction in the bulk, and the analytic results at the edge of the Lanczos spectra using the method of moments.}

\begin{document}

\maketitle
\flushbottom

\section{Introduction}

Tridiagonalization is an important technique in numerical linear algebra that transforms a given matrix into a tridiagonal form, where all non-zero elements are confined to the main diagonal and the primary off-diagonals \cite{viswanath1994recursion}. This transformation simplifies many matrix computations, such as solving eigenvalue problems and performing matrix factorizations. In Hamiltonian systems, tridiagonalization aids in understanding the quantum dynamics of operator growth \cite{Parker:2018a} and the statistical properties of the system \cite{Balasubramanian:2022tpr}. For a Hermitian Hamiltonian, this is typically achieved using the Lanczos algorithm \cite{lanczos1950iteration} or Householder reflections \cite{householder}. The tridiagonal elements, known as Lanczos coefficients are known to efficiently control the dynamics of the system \cite{Nandy:2024htc}. In many contexts, such as the study of orthogonal polynomials, these elements are referred to as recursion coefficients, as they recursively relate to the sequence of orthogonal polynomials \cite{viswanath1994recursion}. This immediately raises an important question about the relationship between the eigenvalues and the Lanczos coefficients. While this might seem a simple question, the answer is often non-trivial. However, in many cases, especially within the context of random matrix theory (RMT), a direct one-to-one correspondence between eigenvalues and Lanczos coefficients may not be necessary. Also, the Lanczos coefficients are not unique; they depend on the chosen initial state which is fed to the Lanczos algorithm. Thus, it can be more insightful to consider statistical questions: Are there correlations between the distribution of eigenvalues, such as the density of states (DOS), and the statistical properties of the Lanczos coefficients?

As it turns out, the answer is affirmative and addressed in \cite{Balasubramanian:2022dnj}. Given the eigenvalues $E_i$ of a Hermitian random matrix, the average DOS $\rho(E)$ is related to the average or mean Lanczos coefficients by 
\begin{align}
        \rho (E) = \int_{0}^1 dx \,\frac{\Theta(4 \mathsf{b}(x)^2 - (E - \mathsf{a}(x))^2)}{\pi \sqrt{4 \mathsf{b}(x)^2 - (E - \mathsf{a}(x))^2}}\,, \label{masrel}
\end{align}
where $\Theta(z)$ is the Heaviside Theta function and $\mathsf{a}(x)$ and $\mathsf{b}(x)$ are the mean Lanczos coefficients, averaged over ensembles. Here $x = n/d \in  [0,1]$, with $n \leq d$ as a rescaled parameter for the coefficients $\{\mathsf{a}_n, \mathsf{b}_n\}$, and $d$ is the dimension of the Krylov space which is upper bounded by the dimension of the Hilbert space or the matrix size. The derivation of \eqref{masrel} can be found in \cite{Balasubramanian:2022dnj}. A similar relationship also holds for the modular spectrum \cite{Caputa:2023vyr}. We will primarily use the notation of \cite{Nandy:2024htc}, to distinguish the Lanczos coefficients derived from the Liouvillian for operator growth \cite{Parker:2018a}. Notably, Eq.\,\eqref{masrel} is valid in the large $d$ limit and in the bulk of the spectrum and does not contain information about the edge $n \sim o(1)$ and the tail $d-n \sim o(1)$ of the spectrum. Note that by ``spectrum'', we are referring to the Lanczos spectrum, not the spectrum of eigenvalues. Although inspired by the seminal work by Dumitriu and Edelman \cite{Dumitriu:2002beta} in RMT, \eqref{masrel} applies to any generic system (chaotic or integrable) with having continuous large $d$ limit, with or without the RMT description \cite{Balasubramanian:2022dnj}.

The Sachdev-Ye-Kitaev (SYK) model \cite{PhysRevLett.70.3339, Kittu} is a $N$ fermionic interaction model where $p$ fermions interact simultaneously. The Hamiltonian is given by:
\begin{align}
    H =  i^{p/2} \sum_{1 \leq i_1 < i_2 \cdots  i_p \leq N} j_{i_1 \cdots i_p} \, \chi_{i_1} \cdots \chi_{i_p}
    \,, \label{hamr}
\end{align}
where $j_{i_1 \cdots i_p}$ are random parameters drawn from a distribution with zero mean and specified variance. The fermions $\chi_{i_k}$ obey the anti-commutation relation among themselves. The model is analytically tractable in the large $N$ and large, finite $p$ limit, where the large $N$ limit is taken before the large $p$ limit \cite{Maldacena:2016hyu}. It is important to remember that although $p$ is taken as large, it does not scale with the system size (i.e., the total number of fermions $N$). On the other hand, the double-scaled limit is defined when $p$ scales with system size $N$ such that $\lambda = 2p^2/N$ is held fixed while both $p, N \rightarrow \infty$ \cite{Cotler:2016fpe, Garcia-Garcia:2018fns, Berkooz:2018jqr}. The large $p$ SYK model and its connection to Jackiw–Teitelboim (JT) gravity are obtained in the so-called triple-scaled limit $\lambda \rightarrow 0$, focusing on the low-energy spectrum which recovers the Schwarzian DOS, after some saddle-point approximations \cite{Berkooz:2018jqr}.

To proceed further, one defines the following parametrization
\begin{align}
    q := e^{- \lambda}\,,~~~~~ \lambda = \lim_{p, N \rightarrow \infty} \frac{2 p^2}{N}\,, \label{ql}
\end{align}
with $\lambda \in (0, \infty)$, i.e., $q \in (1,0)$. In this paper, we will not be concerned with the DSSYK model itself, especially the detailed physics it contains in the triple-scaled limit and its relation to de Sitter space \cite{Susskind:2021esx, Lin:2022rbf}. Also, its algebraic structure \cite{Berkooz:2018jqr, Lin:2023trc}, and path-integral representation \cite{Blommaert:2023opb, Almheiri:2024xtw, Berkooz:2024ofm} are interesting in their own right. A comprehensive review of such physics can be found in \cite{Berkooz:2024lgq}.

On the other hand, our primary focus is on the global average DOS in the Double-Scaled SYK (DSSYK) model. The statement of our problem is as follows. The DOS is given by the $q$-normal distribution, which interpolates between the semicircle law and the normal distribution. Using this as an input for the LHS in \eqref{masrel}, we want to find the mean Lanczos coefficients $\mathsf{a}(x)$ and $\mathsf{b}(x)$. The evaluation of these coefficients requires solving the integral equation \eqref{masrel} which is non-trivial. In the following, we employ three distinct methods to solve it and obtain the mean Lanczos coefficients. 

First, we use a numerical approach by constructing a finite-dimensional Hamiltonian $H_m$ whose DOS mimics the global DOS of the DSSYK model. We randomly select the entries of the Hamiltonian from the DOS, making our Hamiltonian completely integrable by construction (see also \cite{Balasubramanian:2022tpr, Balasubramanian:2022dnj, Balasubramanian:2023kwd, Erdmenger:2023wjg} for similar approaches). We then tridiagonalize this Hamiltonian to obtain the mean Lanczos coefficients. This construction is somewhat limited as it focuses solely on reproducing the DOS, and does not account for any underlying correlations between the eigenvalues. Since the mean Lanczos coefficients depend solely on the global DOS through \eqref{masrel}, we can determine such a Hamiltonian and its tridiagonalization, regardless of whether the Hamiltonian is integrable or chaotic. A similar construction has also been done in the Ising model in both chaotic and integrable limits \cite{Balasubramanian:2023kwd}.

Our second method involves directly solving the integral equation \eqref{masrel} using the semi-analytic approach described in \cite{Balasubramanian:2022dnj}. We outline this method in Sec.\,\ref{secsemiana}. This result is supplemented by a third method that derives an RMT potential for the DOS of the DSSYK model and solves it using saddle-point methods. This potential was initially identified as a matrix model potential for a frustrated spherical model \cite{Cappelli:1997pm}. Recently, it has been revived \cite{Jafferis:2022wez} in the context of Jackiw-Teitelboim (JT) gravity  \cite{JackiwJT, TeitelboimJT} and random matrix models in their double-scaled limit \cite{Saad:2019lba}. The Lanczos coefficients obtained by both methods show excellent agreement with the numerical results in the bulk of the spectrum as the size of the Hamiltonian $H_m$ increases. The bulk Lanczos coefficients are well described by a $q$-deformed logarithm function across the entire parametric regime of $q \in [0,1]$, which marks the \emph{Lanczos descent} \cite{Rabinovici:2020operator, Kar:2021nbm}, going beyond the analysis of \cite{Rabinovici:2023yex}. Edge corrections are considered separately, and we match these by computing the Lanczos coefficients using the moment method \cite{Balasubramanian:2023kwd}. This deformed logarithmic function has also recently been derived in a generic random matrix model (Rosenzweig-Porter model \cite{RPmodel}), which exhibits a transition from ergodic to localized regimes through a fractal phase \cite{Bhattacharjee:2024yxj}.

The motivation for this study from a gravitational perspective is worth highlighting. In the realm of holographic correspondence, JT gravity \cite{JackiwJT, TeitelboimJT} is proposed to share duality with a unitary ensemble, specifically a matrix model in the double-scaled limit \cite{Saad:2019lba}. The full JT gravity partition function with $n$-boundaries, computed by the gravitational path integral, can be expressed as an RMT matrix integral with a specified potential. Interestingly, this path integral is computed on geometries that include higher genus contributions, such as spacetime wormholes (for an excellent review, see \cite{Mertens:2022irh}).  These spacetime wormholes should be distinguished from spatial wormholes like the Einstein-Rosen bridge. The non-trivial geometries of these wormholes emerge as non-perturbative effects of the gravitational path integral, which are believed to control the \emph{Lanczos descent} in the dual RMT \cite{Kar:2021nbm}. Given that the late-time dynamics are governed by the Lanczos descent, it is intriguing to explore how spacetime wormholes in the non-perturbative gravity provide insights into the dynamics of the dual field theory at the boundary quantum theory. The exact form of the Lanczos descent in the DSSYK model that we derive in this paper could help shed light on this duality.

This paper is structured as follows. Section \ref{secDOS} introduces the DOS of the DSSYK model and discusses the Lanczos coefficients in two extreme limits. In Sec.\,\ref{secnum}, we construct the Hamiltonian and obtain the Lanczos coefficients numerically. In the bulk of the spectra, the numerical results are validated using the semi-analytic approach in Sec.\,\ref{secsemiana} and the saddle-point method in Sec.\,\ref{secrmt}. The edge corrections are also addressed in Sec.\,\ref{secsemiana}. We conclude with a summary in Sec.\,\ref{secconc}. \\
\newline
\emph{Note:} During the final stages of work, I became aware of an ongoing work \cite{VijayDSSYK} that has partial overlap with the result of this paper.

\section{Density of States in DSSYK Model} \label{secDOS}

In this paper, we focus on the global DOS in the DSSYK model. A similar structure appears in the $p$-spin glass model \cite{Erdos2014}, Parisi's hypercube model \cite{GParisi_1994, Jia:2020rfn, Berkooz:2023scv} in the orthogonality condition for $q$-Hermite polynomials \cite{ISMAIL1987379, Garcia-Garcia:2017pzl}. The DOS is given by the $q$-normal distribution
\cite{Cotler:2016fpe, Berkooz:2018qkz}:
\begin{align}
    \rho_q(E) = \frac{\sqrt{1-q}} {\pi  \sqrt{1- \frac{1-q}{4}E^2 }}\prod _{k=0}^{\infty }\left(\frac{1-q^{2 k+2}}{1-q^{2 k+1}}\right)\left(1-\frac{(1-q) q^k}{\left(1+q^k\right)^2} E^2 \right)\,.
    \label{denB}
\end{align}
The energy ranges between $E \in \left[-E_m, E_m \right]$, where $E_m = 2/\sqrt{1-q}$, derived from the singularity structure of \eqref{denB}. The DOS is normalized such that
\begin{align}
   \int_{-E_m}^{E_m}  \rho_q(E)\, d E = 1\,. \label{normrho}
\end{align}
For $q \rightarrow 0^{+}$ i.e., $\lambda \rightarrow \infty$ DOS follows the celebrated Wigner semicircle law \cite{wignersemic}. This limit corresponds to $p \gg \sqrt{N}$ in \eqref{ql}. Conversely, for $q \rightarrow 1^{-}$ i.e., $\lambda \rightarrow 0$, the DOS follows the standard normal distribution with zero mean and unit variance, which is analogous to $p \ll \sqrt{N}$. In the SYK model, this limit corresponds to fixing the number of interacting fermions $p$ while taking the total number of fermions $N$ to be large. Thus, for the global DOS, we have
\begin{align}
\rho_q(E) =  \begin{cases} 
   \sqrt{4 - E^2}/(2 \pi) \,,~~~ E \in [-2,2]\,,  & \,~~q \rightarrow 0^{+}\,, \\
   e^{-E^2/2}/(\sqrt{2\pi})  \,,\,~~~ E \in (-\infty, \infty)\,,      & \,~~q \rightarrow 1^{-}\,.\label{eq:DOSq}
  \end{cases} 
 \end{align}
In the latter case, the low-energy spectrum follows the Schwarzian DOS, i.e., $\rho(E) \propto \sinh\small(2\pi \sqrt{E}\small)$ \cite{Cotler:2016fpe, Berkooz:2018qkz}, often referred to as the ``triple-scaled limit'' \cite{Cotler:2016fpe}. We will not discuss it here and refer to the comprehensive review \cite{Berkooz:2024lgq} for more details. Instead, we focus on the global DOS, which is pointwise Gaussian. In both cases in \eqref{eq:DOSq}, one can solve the integral equation \eqref{masrel} to obtain the mean Lanczos coefficients. They are given by
\begin{align}
    &\mathsf{a}_{0^{+}} (x) = 0\,,~~~~ \mathsf{b}_{0^{+}} (x) = \sqrt{1-x}\,\,, ~~~~~~~~ (q = 0^{+})\,, \label{eq:lancq0} \\
    &\mathsf{a}_{1^{-}} (x) = 0\,,~~~~ \mathsf{b}_{1^{-}} (x) = \sqrt{-\frac{1}{2} \ln x}\,, ~~~~~ (q = 1^{-})\,. \label{eq:lancq1}
\end{align}
with $x = n/d$, and $d$ being the system size. The first result is reported in \cite{Kar:2021nbm, Balasubramanian:2022dnj} and the second one is in \cite{Bhattacharjee:2024yxj}.\footnote{I also got to know about the same result by an ongoing work by D. Chakraborty, A. Dymarsky, and R. Ismail. } Numerically, one can take a matrix from Gaussian Unitary Ensemble (GUE) of dimension $d$ and tridiagonalize it to obtain \eqref{eq:lancq0}. However, the eigenvalue distribution will depend on the variance. Hence, we need to rescale the variance properly. To get \eqref{eq:lancq0}, one usually takes the variance of off-diagonal entries chosen as $1/d$, which scales inversely to the system size \cite{Balasubramanian:2022dnj}. For the GOE matrix, the variance of off-diagonal entries needs to be taken as $2/d$ to obtain \eqref{eq:lancq0}. Alternatively, one can determine the eigenvalues with unit variance and then scale them appropriately to follow the semicircle law.

In fact, in the context of nuclear physics, the results \eqref{eq:lancq0}-\eqref{eq:lancq1} are previously known \cite{zuker2001canonical} and extensively used to study the spectral properties of heavy nuclei, dating back to the inspiration of Wigner \cite{Wigner1}. In particular, \eqref{eq:lancq1} has been known to be an approximate version of the inverse Binomial function \cite{Bhattacharjee:2024yxj, zuker2001canonical}. It is important to note that these results are valid in the large $d$ limit and in the bulk of the spectrum.

Despite these two limits, the mean value of the Lanczos coefficients for more general distributions remains unknown.  Progress in this direction has been initiated in \cite{Balasubramanian:2023kwd, Bhattacharjee:2024yxj}. This leads us to the following question: What are the mean values of the Lanczos coefficients for $\rho_q(E)$ defined by \eqref{denB} in the intermediate regime of $q \in  (0,1)$? Can we find analytic forms of the coefficients that converge to both limits in \eqref{eq:lancq0}-\eqref{eq:lancq1}?

The answer for the $\mathsf{a}$-type Lanczos coefficients is straightforward. If the $\mathsf{a}$-type coefficients vanish, the density of states is even, which is easily seen from \eqref{masrel}. However, although not obvious, the reverse statement is also true: if $\rho(E)$ is even then $\mathsf{a}$-type coefficients must vanish \cite{Balasubramanian:2022dnj}. Applying to our example, since the $q$-normal distribution \eqref{denB} is even in $E$, all $\mathsf{a}$-type Lanczos coefficients will vanish leaving only non-zero $\mathsf{b}$-type coefficients. Therefore, our interest lies in finding the mean value of the coefficients $\mathsf{b}(x)$ only. In the following sections, we provide the solution from three different approaches.

\section{Construction of a Hamiltonian: Numerical Approach} \label{secnum}

Our first approach is entirely numerical. For each $q \in  [0,1]$, we construct a model Hamiltonian $H_{m}$ of dimension $d$ such that the eigenvalue DOS of $H_{m}$ represents the DOS \eqref{denB}, with corrections of order $o(1/d)$. Essentially, as $d$ increases, the DOS of the model Hamiltonian more closely approximates the DOS \eqref{denB}. One way to numerically simulate the DOS is to draw independent and uncorrelated random numbers from the distribution \eqref{denB}, which correspond to the eigenvalues of the model Hamiltonian $H_m$. Correspondingly, the average over ensembles is always implied, even if not explicitly stated.

\begin{figure}[t]
   \centering
\includegraphics[width=0.9\textwidth]{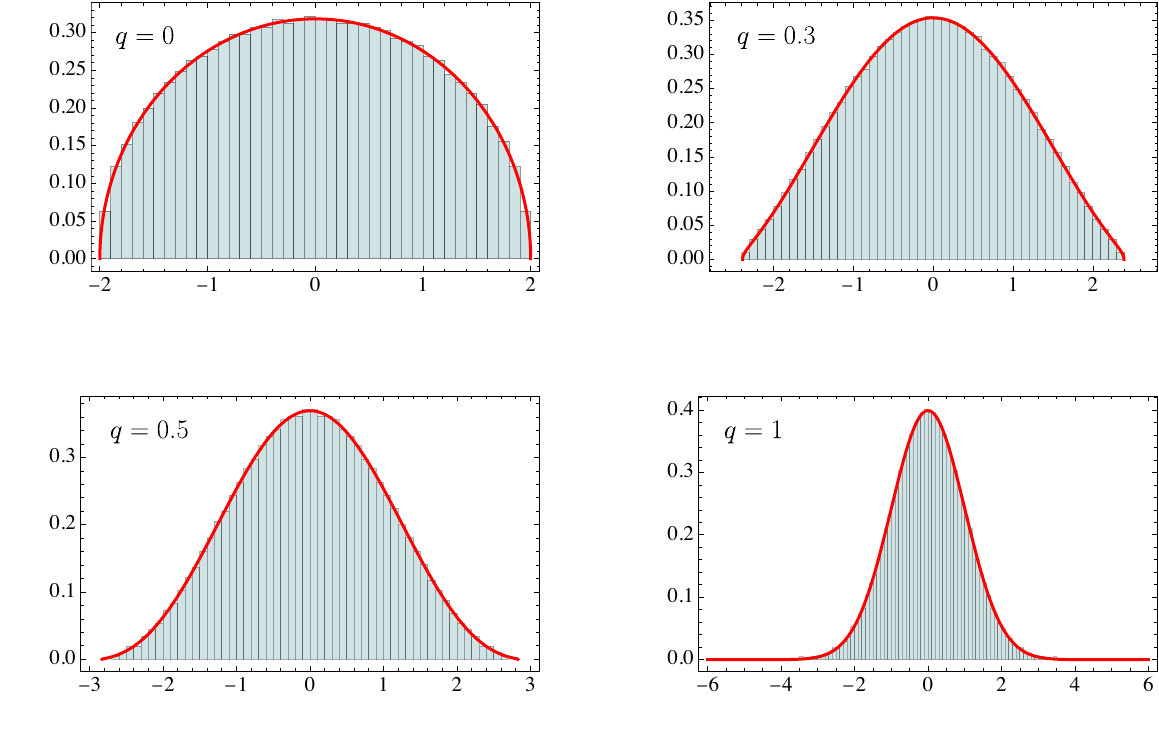}
\caption{The histogram shows the eigenvalue distribution for the model Hamiltonian $H_m$ at various values of $q$. The red line represents the $q$-normal DOS as defined in \eqref{denB}. For $q=0$ and $q=1$, it corresponds to the semicircle law and the normal distribution respectively. For intermediate values of $q$, the infinite product is truncated at $k = 100$. The Hamiltonian $H_m$ has a size of $d = 2048$ and the results are averaged over $300$ ensembles.} \label{fig:allhist}
\end{figure}

Figure \ref{fig:allhist} shows the histogram of the eigenvalue distribution for different values of $q = \{0, 0.3, 0.5, 1\}$ with $d = 2048$ and $300$ ensemble averages. The results show good agreement with the analytic result \eqref{denB} for all values of $q$. Specifically, the two extreme cases, $q=0$ and $q = 1$ correspond to the semicircle law and the normal distribution, respectively, where \eqref{denB} can be explicitly evaluated, see Eq.\eqref{eq:DOSq}. For intermediate values of $q$, the infinite product in \eqref{denB} cannot be explicitly evaluated. Thus we truncate the product at $k = 100$, which provides a reasonable approximation to the exact analytic result.

\begin{figure}[t]
   \centering
\includegraphics[width=\textwidth]{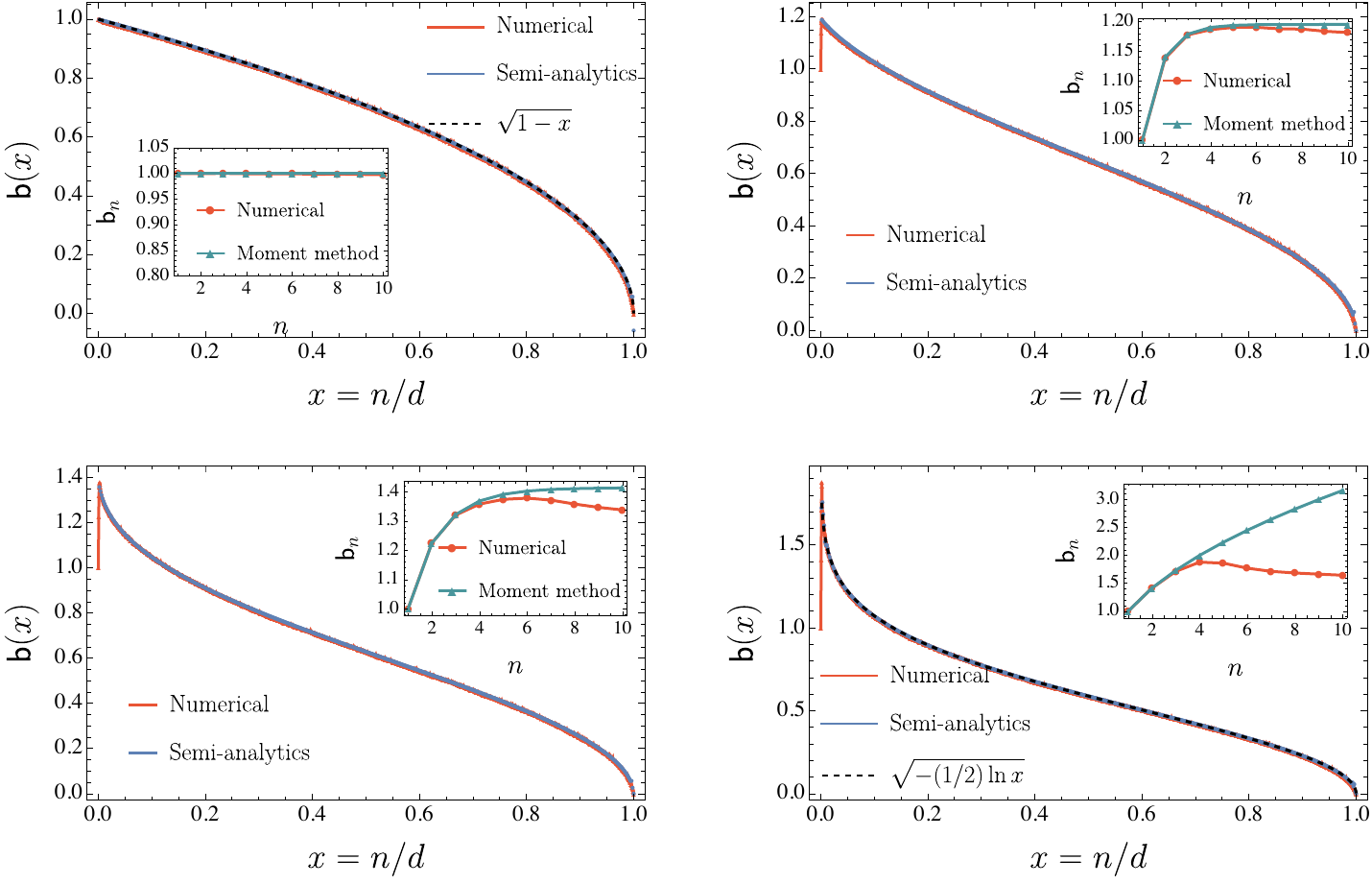}
\caption{The mean value of the Lanczos coefficients for the model Hamiltonian $H_m$ for $q = 0$ (top, left), $q = 0.3$ (top, right), $q = 0.5$ (bottom, left), and $q = 1$ (bottom, right). The red line represents the numerical results, starting from the TFD state and applying the Hessenberg decomposition. The system size is $d = 2048$, and we take averages over $300$ ensembles. The blue line shows the semi-analytic results obtained by solving the integral equation \eqref{masrel}. Here $\mathsf{b}_n$ with $n \sim o(1)$ and $d-n \sim o(1)$ regions are referred to as the ``edge'', and ``tail'' respectively. The ``bulk'' represents the rest of the spectrum, which is highlighted in blue. Both sets of results exhibit excellent agreement in the bulk of the Lanczos spectrum. Insets highlight the edge results, where the teal-colored points are derived using the moment method from the DOS as per \eqref{momden}. The analytic expression for $\mathsf{b}_n$ is known and given by \eqref{bnana} which asymptotes to $1/\sqrt{1-q}$.} \label{fig:bnfullq}
\end{figure}

Next, we proceed to tridiagonalize the Hamiltonian. We start with a diagonal Hamiltonian and rotate it using a random orthogonal matrix drawn from a circular real ensemble (CRE) to preserve the eigenvalues. The resultant Hamiltonian is then tridiagonalized using either the Lanczos algorithm or Householder reflections to obtain the Lanczos coefficients. We begin with the infinite-temperature thermofield double (TFD) state \cite{Maldacena:2001kr} as our initial state, although any generic random state can also be used. We will briefly comment on the choice of the initial state later.

The tridiagonal representation retains the same eigenvalues as the original Hamiltonian. This process is repeated multiple times, and we perform an ensemble average over the Lanczos coefficients. Note that the initial states are related to the orthogonal transformations on the diagonal Hamiltonian, which also modifies the eigenvectors. Since our Hamiltonian is constructed with random elements and we are concerned with the statistical distribution rather than specific elements, these modifications do not affect our conclusions.

Figure \ref{fig:bnfullq} (red lines) illustrates the behavior of the Lanczos coefficients for four different values of $q$, corresponding to the histograms shown in Fig.\,\ref{fig:allhist}. The $x$ axis is appropriately scaled. We observe that the $q=0$ and $q=1$, the Lanczos coefficients are properly given by \eqref{eq:lancq0} and \eqref{eq:lancq0} respectively, which corresponds to the semicircle and the normal distribution respectively. The Lanczos spectrum for intermediate values of $q$ is undetermined. Note that this method provides the entire spectrum, including the ``edge'', ``bulk'' and ``tail'', appropriately mentioned in Fig.\,\ref{fig:bnfullq}.

It is important to note that our Hamiltonian $H_m$ mimics the DSSYK model only at the level of the DOS and does not account for other correlations. The Hamiltonian is constructed with elements independently drawn from the DOS, making it integrable. To see this, we can compute the Krylov (state) complexity $ K_S(t) = \sum_n n\, |\uppsi_n|^2$, as the average position on the Krylov chain, where $\uppsi_n$ satisfies the three-term recurrence relation. This notion is inspired by the Krylov complexity in operator space, first introduced in Ref.\,\cite{Parker:2018a} within the context of many-body systems and the SYK model. It was subsequently extended to quantum field theories \cite{Dymarsky:2021bjq, Avdoshkin:2022xuw, Camargo:2022rnt}. For a comprehensive review, see \cite{Nandy:2024htc}. The complexity depends on the choice of the initial state and we choose a particular state: the infinite-temperature TFD state \cite{Maldacena:2001kr}, namely, the uniform superposition of the eigenstates of the Hamiltonian. We note that the mean Lanczos coefficients in the bulk are independent of the initial choice of the TFD state. The same mean coefficients can be obtained when selecting any random state. However, at the edge, the result with the TFD state fits slightly better than any random state. This is because the moments (see \eqref{momden} later) are evaluated using the trace operator, where all the states contribute. However, such differences become insignificant in the large $d$ limit. Since the model Hamiltonian $H_m$ is integrable, the Krylov state complexity does not exhibit any peak. However, the exact mean value of the Lanczos coefficients in the $q \rightarrow 0$ limit (with less variance) can also be obtained from a correlated random matrix such as the GUE. In such scenarios, the complexity does exhibit a peak. Therefore, the mean Lanczos coefficients are insufficient to determine the presence of eigenvalue correlations \cite{Balasubramanian:2023kwd, Erdmenger:2023wjg}, while this controls the saturation value. This further suggests that the peak of Krylov complexity incorporates the non-trivial correlation of the Lanczos spectrum. Note that, the complexity, computed this way involves the state picture, thus it should not be directly compared with the result in the operator picture \cite{Bhattacharjee:2022ave, Aguilar-Gutierrez:2024nau}.

\section{Derivation from Semi-analytic Approach} \label{secsemiana}

The numerical results in the previous section are valid for the entire spectrum of Lanczos coefficients, including the edge and the tail of the spectra. In this section, we directly solve the integral equation \eqref{masrel} for each value of $q$ to determine the mean Lanczos coefficients from the DOS \eqref{denB}. As previously mentioned, this approach yields the Lanczos coefficients within the bulk of the spectra, excluding coefficients where $n \sim o(1)$. The integral equation \eqref{masrel} is then solved by the bisection method. Notice that the support of $E$ in the integral \eqref{denB} ranges from $E_{\mathrm{left}}(x) = a(x) - 2 b(x)$ to $E_{\mathrm{right}}(x) = a(x) + 2 b(x)$. Assuming that the integral shrinks as $x$ increases, the cumulative DOS is given by \cite{Balasubramanian:2022dnj}
\begin{align}
    P(E) = \int_{E_{\mathrm{min}}}^E \rho(E') dE' \approx \int_{0}^1 dx \int_{E_{\mathrm{left} }(x)}^E dE'
 \,\frac{\Theta(4 \mathsf{b}(x)^2 - (E' - \mathsf{a}(x))^2)}{\pi \sqrt{4 \mathsf{b}(x)^2 - (E' - \mathsf{a}(x))^2}}\,.
\end{align}
Defining the integral $P_c(z) = \int_{-2}^z \frac{dz'}{\pi\sqrt{4-z'^2}}$, we rewrite the above equation as
\begin{align}
    P(E) = \int_{0}^1 dx P_c \bigg(4\frac{E - E_{\mathrm{left}} (x)}{E_{\mathrm{right}} (x)-E_{\mathrm{left}} (x)}-2\bigg)\,,
\end{align}
where we have used the substitution $E = b z' - a$. A schematic description of the above prescription is given by the following algorithm, which is also taken from \cite{Balasubramanian:2022dnj}:

\begin{algorithm}[H]
\caption{Solving the integral equation \eqref{denB}}
\begin{algorithmic}[1]
\State $E_{\mathrm{right}}(0) \gets E_{\mathrm{max}}$.
\State $E_{\mathrm{left}}(0) \gets E_{\mathrm{min}}$.
\State for $n \in 1 : N$ do
\State Set $E_{\mathrm{left}}^{(n/N)}$ as the the lowest solution $E > E_{\mathrm{left}}^{(n-1)/N}$ of
\begin{align}
P(E) = \frac{1}{N} \sum_{i=0}^{n-1} P_c \left( \frac{4(E - E_{\mathrm{left}}^{i/N})}{E_{\mathrm{right}}^{i/N} - E_{\mathrm{left}}^{i/N}} - 2 \right)\,.
\end{align}
\State Set $E_{\mathrm{right}}^{(n/N)}$ as the highest solution $E < E_{\mathrm{right}}^{(n-1)/N}$ of
\begin{align}
1 - P(E) = \frac{1}{N} \sum_{i=0}^{n-1} \left[ 1 - P_c \left( \frac{4(E - E_{\mathrm{left}}^{i/N})}{E_{\mathrm{right}}^{i/N} - E_{\mathrm{left}}^{i/N}} - 2 \right)\right]\,.
\end{align}
\State $\mathrm{end~for}$
\State $a \gets \frac{E_{\mathrm{left}} + E_{\mathrm{right}}}{2}$.
\State $b \gets E_{\mathrm{right}}$.
\end{algorithmic}
\end{algorithm}

\begin{figure}[t]
   \centering
\includegraphics[width=0.53\textwidth]{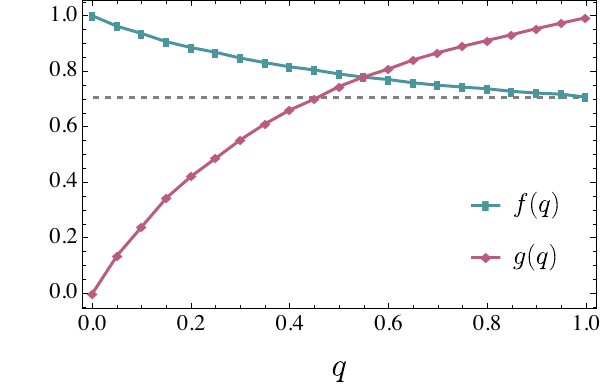}
\caption{The variation of $f(q)$ and $g(q)$ within the range $q \in [0,1]$. At $q=1$, the function $f(q)$ reaches a value $=1/\sqrt{2} \approx 0.707$, as indicated by the black dashed line. This value is consistent with the Lanczos coefficients for the normal distribution as given in \eqref{eq:lancq1}.} \label{fig:fgplot}
\end{figure}

Figure \ref{fig:bnfullq} illustrates the results (depicted in blue) for four specific values (same values as the histogram plots and numerical results are done in Sec.\,\ref{secnum}) of $q$ including the values corresponding to the semicircle and the normal distribution. The results in the bulk spectrum align perfectly with the numerical tridiagonalization obtained in the previous section. Except for the two extremes regimes of $q$, the analytic form of the coefficients for intermediate values of $q$ in the bulk spectrum remains undetermined. To address this, we propose the following form in the bulk:
\begin{align}
    \mathsf{b}(x) = f(q) \sqrt{-\ln_{g(q)} x}\,, \label{mathprop}
\end{align}
Here $f(q)$ and $g(q)$ are two continuous functions of $q$, which parameterized $\mathsf{b}(x)$. Additionally, $\ln_{\mathsf{q}} x$ is known as the $\mathsf{q}$-deformed logarithm,\footnote{We denote $\mathsf{q}$ to define the deformed logarithm which equals $g(q)$ for our case.} defined as
\begin{align}
    \ln_{\mathsf{q}} x = \frac{x^{1-\mathsf{q}}-1}{1-\mathsf{q}}\,,~~~~~~~ \lim_{\mathsf{q} \rightarrow 0} = x-1\,,~~~~~~~ \lim_{\mathsf{q} \rightarrow 1} \ln_{\mathsf{q}} x = \ln x\,. \label{qdlog}
\end{align}
The limiting expressions \eqref{eq:lancq1} suggests $f(0) = 1$,  $f(1) = 1/\sqrt{2}$, and $g(0) = 0$, $g(1) = 1$. The entire profile of $f(q)$ and $g(q)$ can be found by fitting the proposed form \eqref{mathprop} to the numerical result for each $q \in [0,1]$. The resulting profiles are shown in Fig.\,\ref{fig:fgplot}. The function $f(q)$ decreases with $q$, starting at unity for  $q = 0$ and reaching $1/\sqrt{2} \approx 0.707$ at $q = 1$. On the other hand, $g(q)$ is an increasing function of $q$; it vanishes for $q = 0$ while reaching unity at $q = 1$. This limiting behavior is consistent with the structure of the $\mathsf{q}$-deformed logarithm \eqref{qdlog}. The structure of \eqref{mathprop} is inspired by a recent study \cite{Bhattacharjee:2024yxj} on Krylov space dynamics in the Rosenzweig-Porter matrix model \cite{RPmodel}. 

It is interesting to obtain the qualitative behavior of $g(q)$, in particular. A non-linear fitting suggests the following behavior:
\begin{align}
    g(q) = u \,q^v\,, ~~~~ u = 1.03\,,~~~ v = 0.54\,,
\end{align}
which differs slightly from the square-root growth. The system size is taken as $d = 2048$ over $300$ samples. We find that the coefficient and the exponent weakly depend on $d$, hence, in the large $d$ limit, the square-root growth cannot be ruled out. We also observe that at a particular $q$, the functions $f(q)$ and $g(q)$ become equal, which is around $q \approx 0.57$. However, this value does not hold any substantial significance in our current analysis.

\begin{figure}[t]
   \centering
\includegraphics[width=0.8\textwidth]{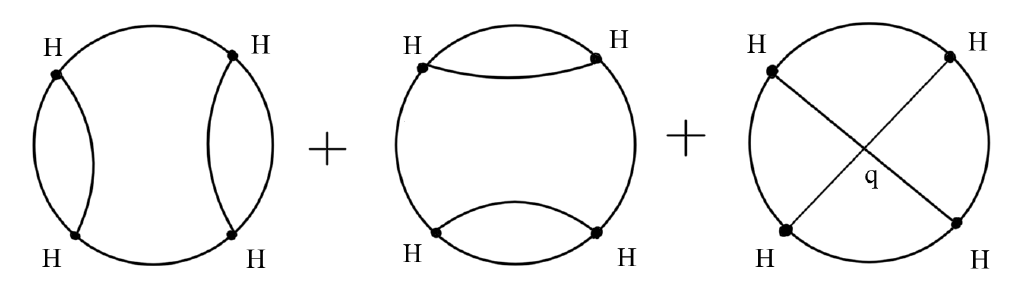}
\caption{The chord diagrams for evaluating $m_4 =  \langle \mathrm{Tr}(H^4) \rangle_J$ consist of three diagrams in total. Each diagram features four nodes, representing the four insertions of the Hamiltonian, with chords connecting them in various configurations. The left and middle diagrams have no chord crossings, thus they each evaluate to unity. The right diagram has a single crossing, which evaluates to $q$. Therefore, the total contribution is $m_4 = 1 +  1 + q = 2+q$.} \label{fig:chord}
\end{figure}

At the edges, particularly where $n \sim o(1)$, there are significant deviations from the numerical results, as can be seen from Fig.\,\ref{fig:bnfullq}. This is expected, since \eqref{masrel} does not hold in this limit. To address this regime, we employ the analytic moment method. We first compute the moments of the distribution
\begin{align}
    m_k = \langle \mathrm{Tr}(H^k) \rangle_J =
   \int_{-E_m}^{E_m}  E^k \rho_q(E)\, d E\,, \label{momden}
\end{align}
averaged over the random couplings and $E_m = 2/\sqrt{1-q}$. They are the moments of the DSSYK Hamiltonian. The normalization condition \eqref{normrho} corresponds to $m_0 = 1$. Further, since $\rho_q(E)$ is an even function of $E$, the odd moments vanish. The even-moments are given by \cite{Berkooz:2018qkz}
\begin{align}
    m_{2k} = \frac{1}{(1-q)^k} \sum_{j = -k}^k (-1)^j q^{j(j-1)/2} \binom{2k}{k+j}\,. \label{momex}
\end{align}
As an example, $m_2 = 1$, $m_4 = 2 + q$, $m_6 = 5 + 6q + 3q^2 + q^3$, and so on. They can also be evaluated using the chord diagram techniques \cite{Berkooz:2018jqr}. See Fig.\,\ref{fig:chord} as an example. In the two extreme limits, they are given by \cite{Erdos2014}
\begin{align}
    \lim_{q \rightarrow 0} m_{2n} = \frac{(2n)!}{n! (n+1)!}\,,~~~~~~~ \lim_{q \rightarrow 1} m_{2n} = \frac{(2n)!}{2^n n!}\,.
\end{align} 
The Lanczos coefficients are given by solving the iterative algorithm 
\cite{viswanath1994recursion, Parker:2018a}
\begin{align}
    \mathsf{b}_n &= \sqrt{Q_{2n}^{(n)}}\,, ~~~~ Q_{2k}^{(m)} = \frac{Q_{2k}^{(m-1)}}{\mathsf{b}_{m-1}^2} - \frac{Q_{2k-2}^{(m-2)}}{\mathsf{b}_{m-2}^2}\,, \nonumber \\
    Q_{2k}^{(0)} &= m_{2 k}\,, ~~~~ \mathsf{b}_{-1} = \mathsf{b}_{0} := 1\,, ~~~~ Q_{2k}^{(-1)} := 0\,. \label{mombn}
\end{align}
Feeding the moments \eqref{momex}, we obtain the Lanczos coefficients as\footnote{Ref.\,\cite{Rabinovici:2023yex} has obtained this form with an additional factor of $J/\sqrt{\lambda}$ due to a different choice of the variance. Also, the expression is valid for $n \ll d$, for large $d$ systems.} \cite{Rabinovici:2023yex}
\begin{align}
    \mathsf{b}_n = \sqrt{\frac{1-q^n}{1-q}}\,. \label{bnana}
\end{align}
The coefficients $\mathsf{a}_n$ vanish, due to the vanishing odd moments. They can also be directly read from the corresponding transfer matrix construction using chord diagram techniques \cite{Berkooz:2018jqr}. Since they are obtained using the Hamiltonian moments, their growth is \emph{not related} to the operator growth hypothesis \cite{Parker:2018a}. In the extreme limits, they are given by
\begin{align}
    \lim_{q \rightarrow 0} \mathsf{b}_{n} = 1\,~~ (\mathrm{semicircle})\,, ~~~~~~ \lim_{q \rightarrow 1} \mathsf{b}_{n} = \sqrt{n}\, ~~(\mathrm{normal})\,.
\end{align}
We compare them with the edge Lanczos coefficients obtained numerically. The insets of Fig.\,\ref{fig:bnfullq} show such a comparison for the four values of $q$. The analytic result asymptotes to $1/\sqrt{1-q}$. We see that the numerical results fairly match with the analytic results for small $q$. The match gets better as $d$ increases. This is because \eqref{bnana} corresponds to the infinite-size system while we only match them with finite-size systems. However, around $q = 1$, the match is poor. The analytic result approaches $\sim \sqrt{n}$, which requires a larger matrix in numerical computation.  Notice that, for $x \sim o(1)$, the bulk Lanczos coefficients \eqref{mathprop} become $\mathsf{b} \sim f(q)/\sqrt{1-g(q)}$. This can be appropriately \emph{padded} with the asymptotic edge result $\sim 1/\sqrt{1-q}$ (from \eqref{bnana} with large $n$) within a specific range of $n$, using the values of $f(q)$ and $g(q)$ obtained from Fig.\,\ref{fig:fgplot}. This marks the beginning of \emph{Lanczos descent}. A similar descent has also been observed in the operator picture \cite{Rabinovici:2020operator, Kar:2021nbm}.

\section{Construction via Random Matrix Theory: Saddle-point Method} \label{secrmt}

We now provide an alternative way to obtain the mean Lanczos coefficients in the bulk, using the saddle-point method in the large $d$ limit through the RMT construction \cite{Balasubramanian:2022dnj}. This approach essentially investigates whether a tridiagonal ensemble can be derived for any arbitrary RMT potential. The motivation behind this is straightforward. It is well known that the joint probability distribution of the eigenvalues is given by \cite{RMTbook}
\begin{align}
    p(\lambda_1, \ldots, \lambda_d) = Z_{\upbeta, d} \prod_{i < j} |\lambda_i - \lambda_j|^{\upbeta} e^{-\frac{\upbeta d}{4} \mathrm{Tr}V(H)} \,, \label{diagdist}
\end{align}
where $Z_{\upbeta, d}$ is a normalization factor, $d$ is the dimension of the matrix, and $V(H)$ is any arbitrary potential such that the trace remains invariant under the unitary transformation on the Hamiltonian. The multiplicative factor is known as the Vandermonde determinant, and $\upbeta$ is Dyson's index, which counts the number of real components in the matrix elements; $\upbeta = 1,2,4$ for the Gaussian Orthogonal Ensemble (GOE), Gaussian Unitary Ensemble (GUE), and Gaussian Symplectic Ensemble (GSE), respectively. Equation \eqref{diagdist} represents the probability distribution of the diagonal elements (i.e., the eigenvalues). The potential is related to the DOS through the principal value (P.V.) integral \cite{RMTbook}
\begin{align}
    \frac{1}{4} V'(E) = \mathrm{P.V} \int dE' \, \frac{\rho(E')}{E-E'}\,. \label{pv}
\end{align}
However, the same matrix can also be tridiagonalized. It is then reasonable to study the probability distribution of the tridiagonal elements, namely the Lanczos coefficients. This procedure was established by the seminal work of Dumitriu and Edelman \cite{Dumitriu:2002beta}. The result is given by \cite{Balasubramanian:2022tpr}:
\begin{align}
    p(a_0, \cdots, a_{d-1}, b_1, \cdots, b_{d-1}) = \left(\prod_{n=1}^{d-1} b_n^{(d-n) \upbeta -1}\right) e^{-\frac{\upbeta d}{4} \mathrm{Tr}V(H)}\,.
\end{align}
Comparing this to the diagonal distribution \eqref{diagdist}, we see that this relates the Lanczos coefficients to the potential. Through \eqref{pv}, one can relate the DOS to the mean Lanczos coefficients, as given in \eqref{masrel}. We refer to \cite{Balasubramanian:2022tpr} for technical details.

We aim to seek an effective action for the Lanczos coefficients in the large $d$ limit such that $S_{\mathrm{eff}} = \ln p(a_0, \cdots, a_{d-1}, b_1, \cdots, b_{d-1})$. After some manipulation, one obtains \cite{Balasubramanian:2022dnj} 
\begin{align}
    \frac{S_{\mathrm{eff}}}{\upbeta d^2} = \int dx (1-x) \ln \mathsf{b}(x) - \frac{1}{4} \int dx \int dE \frac{V(E)}{\pi \sqrt{4 \mathsf{b}(x)^2 - (E-\mathsf{a}(x))^2}}\,,
\end{align}
Here, $\upbeta$ is the Dyson index. By extremizing the above action, we get
\begin{align}
    \mathsf{b}(x) \frac{\partial}{\partial \mathsf{b}(x)}\left(  \int dE \frac{V(E)}{\pi \sqrt{4 \mathsf{b}(x)^2 - (E-\mathsf{a}(x))^2}}\right) &=  4(1-x)\,, \label{ppot1}\\
    \frac{\partial}{\partial \mathsf{a}(x)}\left(  \int dE \frac{V(E)}{\pi \sqrt{4 \mathsf{b}(x)^2 - (E-\mathsf{a}(x))^2}}\right) &= 0\,, \label{ppot2}
\end{align}
where the integration limits are $E = \mathsf{a}(x) \pm 2 \mathsf{b}(x)$. In other words, the mean value of the Lanczos coefficients extremizes the action. This equation is valid for any generic potential and relates the RMT potential to the mean Lanczos coefficients. However, for polynomial potential $V(E) = \sum_n w_n E^n$, these equations can be simplified. Consider the integral (writing $\mathsf{a} = \mathsf{a}(x)$ and $\mathsf{b} = \mathsf{b}(x)$ for brevity)
\begin{align}
    \int_{\mathsf{a}-2 \mathsf{b}}^{\mathsf{a}+2 \mathsf{b}} dE \frac{V(E)}{\pi \sqrt{4 \mathsf{b}^2 - (E-\mathsf{a})^2}} &= \sum_n \frac{w_n}{\mathsf{b}} \int_{\mathsf{a}-2 \mathsf{b}}^{\mathsf{a}+2 \mathsf{b}} dE \frac{E^n}{\pi \sqrt{4 - \big(\frac{E-\mathsf{a}}{\mathsf{b}}\big)^2}} \nonumber \\
    &= \sum_n w_n \int_{-2}^{2} \frac{(\mathsf{b} y + \mathsf{a})^n}{\sqrt{4-y^2}} ~~~~~~(E = \mathsf{b} y + \mathsf{a}) \nonumber \\
    &= \sum_n w_n \,\binom{n}{m} \mathsf{b}^m \, \mathsf{a}^{n-m} \int_{-2}^{2} \frac{y^m}{\sqrt{4-y^2}} \nonumber \\
    &= \sum_n w_n  \mathsf{b}^m \, \mathsf{a}^{n-m} \,\binom{n}{m} \binom{m}{m/2}
\end{align}
where we have used the integral identity $\int_{-2}^{2} \frac{y^m}{\sqrt{4-y^2}} = \binom{m}{m/2}$ \cite{Balasubramanian:2022dnj}. Hence, after taking the derivative with respect to $\mathsf{a}$ and $\mathsf{b}$, \eqref{ppot1}-\eqref{ppot2} simplifies to a set of algebraic equation:
\begin{align}
    \sum_n w_n \sum_m m\,\mathsf{a}^{n-m} \mathsf{b}^{m} \,\binom{n}{m} \binom{m}{m/2} &= 4(1-x)\,, \label{poleq01} \\
    \sum_n w_n \sum_m (n-m) \,\mathsf{a}^{n-m-1} \mathsf{b}^{m} \,\binom{n}{m} \binom{m}{m/2} &= 0\,. \label{poleq02}
\end{align}
These coupled equations provide the required mean Lanczos coefficients.\footnote{We note that there is a typo in the first identity of Eqn.\,(89) in Ref.\,\cite{Balasubramanian:2022dnj}, which has been corrected in \eqref{poleq01}-\eqref{poleq02}.}

For our case, the primary challenge is to obtain a potential that reproduces the DOS \eqref{denB} of the DSSYK model through the principal value (P.V) integral \eqref{pv}.
Numerically such potential can be obtained by stretching the spectrum \cite{Balasubramanian:2023kwd}. Interestingly, an analytical form of this potential is known:
\begin{align}
    V_q'(E) &= 4\sqrt{1-q} \sum_{n=0}^{\infty} (-1)^n q^{n(n+1)/2} \cosh[(2n+1) \chi]\,, ~~ E= \frac{2 \cosh \chi}{\sqrt{1-q}}\,,\nonumber \\
    &= 2 \sqrt{1-q} \sum _{n=0}^{\infty} (-1)^{n-1} q^{\frac{n^2}{2}} \left(q^{\frac{n}{2}}+q^{-\frac{n}{2}}\right) U_{2 n-1}\left(\frac{E}{2} \sqrt{1-q}\right) \,.
\end{align}
where the prime indicates the derivative with respect to $E$, and $U_{n}(x)$ is the Chebyshev polynomial of the second kind. The first equation is derived in the frustrated Spherical Model \cite{Cappelli:1997pm} in the planar limit \cite{Brezin1978}, while the second one is obtained using the genus expansion in the matrix model for DSSYK model \cite{Jafferis:2022wez, Okuyama:2023kdo}.
One can check, at least numerically, that these two forms are completely equivalent. However, for our case, the second form is more useful since, using the identity $dT_n(x)/dx = n U_{n-1}(x)$, one can integrate it to obtain the exact form of the potential \cite{Jafferis:2022wez}
\begin{align}
    V_q(E) = 2\sum _{n=1}^{\infty} \frac{(-1)^{n-1}}{n} q^{\frac{n^2}{2}} \left(q^{\frac{n}{2}}+q^{-\frac{n}{2}}\right) T_{2 n}\left(\frac{\sqrt{1-q}}{2} E\right)\,.
\end{align}
where $T_{n}(x)$ is the Chebyshev polynomial of the first kind. This, of course, corresponds to a non-Gaussian RMT, and the Gaussian result with $V(E) = E^2 \,+\, \mathrm{constant}$ is recovered in the limit of $q \rightarrow 0$.

To apply the formulas \eqref{poleq01}-\eqref{poleq02}, we intend to express this potential in a polynomial of $E$, for any $q$. Numerically this implies terminating the series (ensuring the convergence) up to some finite and large $M$. Additionally, $T_{2n}(-x) = T_{2n}(x)$ which implies that the even powers of $E$ will be present. Hence, we get
\begin{align}
    V_q(E) \approx \sum_{n = 1}^M w_{2n}^M (q)  E^{2n}\,, \label{pot}
\end{align}
where $w_{2n}^M (q)$ are the expansion coefficients. They depend on $M$, i.e., up to the number of terms we truncate the expansion. Here, the $\approx$ symbol indicates that we indeed perform such truncation ensuring the convergence, such that the contribution of the $(M+1)$-th term is less and does not alter the result.
This is possible because $0\leq q \leq 1$. 
Thus, the Lanczos coefficients satisfy two sets of coupled equations:
\begin{align}
    \sum_n w_n^M (q) \sum_m m\, \mathsf{a}^{n-m} \mathsf{b}^{m} \binom{n}{m} \binom{m}{m/2} &= 4(1-x)\,, \label{poleq1} \\
    \sum_n w_n^M (q) \sum_m (n-m)\, \mathsf{a}^{n-m-1} \mathsf{b}^{m} \binom{n}{m} \binom{m}{m/2} &= 0\,. \label{poleq2}
\end{align}
Since our potential \eqref{pot} is even in $E$, only the $\mathsf{b}$-type coefficients will be present. This can also be directly seen from \eqref{poleq1}-\eqref{poleq2}, after the expansion. The second equation \eqref{poleq2} provides $\mathsf{a}_n = 0$, while the first equation \eqref{poleq1} provides an polynomial expression for $\mathsf{b}_{n}$:
\begin{align}
    \mathsf{w}_{M} \, \mathsf{b}^M + \mathsf{w}_{M-2} \mathsf{b}^{M-2} + \cdots = 0\,. \label{bnss}
\end{align}
where $\mathsf{w}_{M} = f(\{w_n^M (q)\})$, i.e., some polynomial function of $w_n^M (q)$, whose specific form is not important for us. We numerically solve the polynomial equation, focusing on the real solutions. When multiple real solutions exist, we further restrict to the solution with the minimum value of $\mathsf{b}_n$ \cite{Balasubramanian:2022dnj}.

\begin{figure}[t]
   \centering
\includegraphics[width=1\textwidth]{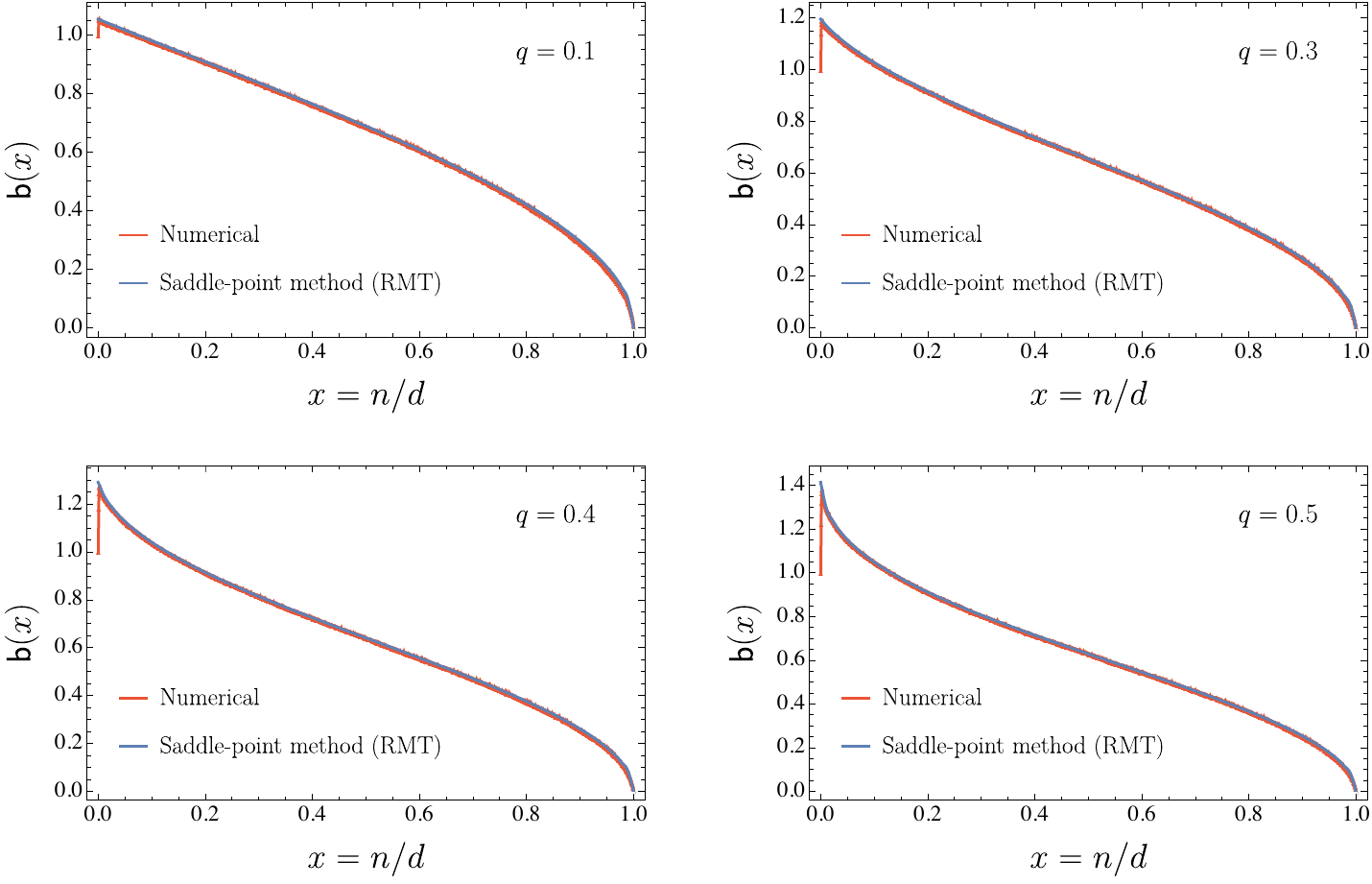}
\caption{The mean value of the Lanczos coefficients obtained from the Hamiltonian $H_m$ (red) is compared with those obtained using the saddle-point method in RMT (blue) for $q = 0$ (top, left), $q = 0.1$ (top, right), $q = 0.3$ (bottom, left), and $q = 0.5$ (bottom, right). The parameters for the integrable (red) case are the same as in Fig.\,\ref{fig:bnfullq}, with $d = 2048$ over $300$ ensembles.} \label{fig:rmtint}
\end{figure}

Figure \ref{fig:rmtint} shows the comparison between the Lanczos coefficients obtained from the model Hamiltonian $H_m$ in Sec.\,\ref{secnum} with those derived using \eqref{bnss}. We focus on $q \leq 0.5$ for four different values of $q$, truncating the expression \eqref{pot} at $M = 100$. The results show excellent agreement in the bulk of the Lanczos spectrum. This is expected, as the mean Lanczos coefficients depend primarily on the global structure of the DOS. Therefore, even in a non-Gaussian RMT scenario,\footnote{While not directly related to our construction, the study of Krylov state complexity in a variant of non-Gaussian RMT has also been explored in \cite{Bhattacharyya:2023grv}.} the Lanczos coefficients can still be expressed in terms of the deformed logarithmic function. For $0.5 < q \leq 1$, higher terms in \eqref{pot} may be required for better convergence, and Pad\'e approximant can be employed. However, this does not alter our conclusion.


\section{Conclusion and Outlook} \label{secconc}

In this paper, we explore the relationship between the density of states (DOS) and the mean Lanczos coefficients of tridiagonal matrices. We focus on examples of DOS that interpolate between the semicircle and normal distributions, such as the SYK model, the spin glass model, and hypercube models in the double-scaled limit. Specifically, for the DSSYK model, the interpolating parameter is given by $q = e^{-2p^2/N}$, where $p$ is the number of interacting fermions and $N$ is the total number of fermions, with the limit $p, N \rightarrow \infty$ and $p^2/N$ held fixed. For each $q$, we construct an ensemble of integrable, effective, finite-dimensional Hamiltonians of dimension $d$ that mimic the DOS of the DSSYK model. Given the finite-dimensional nature of the Hamiltonian, we numerically tridiagonalize it to obtain the mean Lanczos coefficients, starting from the infinite-temperature TFD state \cite{Maldacena:2001kr}. While the mean of Lanczos coefficients is known for the two extreme limits of the DOS, the interpolating regime was previously unknown. We derive the mean value of Lanczos coefficients in terms of a particular $q$-deformed logarithm function, given by \eqref{mathprop}. Our numerical results are validated through both semi-analytical and analytical approaches, including the use of RMT potential via saddle-point methods to examine the bulk and edge Lanczos coefficients. Our focus is solely on solving \eqref{masrel} and does not extend to speculating on any gravitational interpretation of the Krylov complexity. However, this is closely linked with our analysis in Sec.\,\ref{secsemiana} through \eqref{bnana}. In this context, Krylov complexity has been shown to manifest in bulk gravity as the length of a two-sided wormhole in $(1+1)$ dimensions \cite{Rabinovici:2023yex}. The Lanczos coefficients exhibit $\mathsf{b}_n \sim \sqrt{n}$ growth, before reaching a plateau. The descent regime of the Lanczos coefficients is not considered since the DSSYK model in the thermodynamic limit is an infinite-dimensional system.  However, it would be interesting to examine the Lanczos descent (along with the variances) given by \eqref{mathprop} for finite $N$, where non-perturbative effects \cite{Kar:2021nbm}, such as wormholes in the bulk gravitational theory, may influence the late-time behavior of Krylov complexity and the dynamics of the system.

One can also consider a linear combination\footnote{The Krylov complexity (both in operator and state picture) for the linear combination of SYK model are also considered in \cite{Chapman:2024pdw, Baggioli:2024wbz}.} of the chaotic and integrable SYK model where the latter is constructed using the commuting variables in their double-scaled limit \cite{Almheiri:2024xtw, Berkooz:2024ofm}. Especially, due to the normal distribution of the DOS in commuting SYK model \cite{Gao:2023gta}, the Lanczos coefficients will take the form \eqref{eq:lancq1}, with an additional pre-factor in \eqref{mathprop}, as shown in Ref.\,\cite{Bhattacharjee:2024yxj}. We also speculate that our findings in Sec.\,\ref{secrmt} can be obtained by numerically constructing the potential, using methods such as Metropolis sampling \cite{metropolis} or spectrum stretching \cite{Balasubramanian:2023kwd}. It would be interesting to observe how the mean Lanczos coefficients and their covariances describe the late-time dynamics \cite{Balasubramanian:2023kwd} within the single and two-matrix model \cite{Jafferis:2022wez}. We leave these detailed investigations for future work.

\section*{Acknowledgements}

I would like to thank Vijay Balasubramanian, Budhaditya Bhattacharjee, Damian Galante, Yiyang Jia, Javier Magan, Masamichi Miyaji, Kazumi Okuyama, Tanay Pathak, Masaki Tezuka, Qingyue Wu, and Zhuo-Yu Xian for useful discussions. I particularly thank Tanay Pathak for helpful suggestions on numerical computations and Yiyang Jia for insightful feedback that primarily led to Section \ref{secrmt}. Part of this work was presented at the 4th Extreme Universe Annual Meeting at Osaka University. I also acknowledge the hospitality of Osaka University and Kavli IPMU, University of Tokyo during the final stages of this work. The author is supported by the JSPS Grant-in-Aid for Transformative Research Areas (A) ``Extreme Universe'' No.\,21H05190.

\bibliographystyle{JHEP}
\bibliography{refs}

\end{document}